\magnification=1200
\headline{\ifnum\pageno=1 \nopagenumbers
\else \hss\number \pageno \fi}

\footline={\hfil}
\overfullrule=0pt
$$\centerline{\bf PARTON DISTRIBUTIONS IN THE PHOTON}$$\bigskip
\centerline{\bf P. Aurenche and J.-Ph. Guillet}\smallskip
\centerline{Laboratoire de Physique ${\rm Th\acute eorique}$
ENSLAPP\footnote{*}{URA 14-36 du CNRS, ${\rm associ\acute ee}$ ${\rm \grave a}$
l'Ecole Normale ${\rm Sup\acute erieure}$ de Lyon, et au Laboratoire
d'Annecy-le-Vieux de Physique des Particules} - Groupe d'Annecy}\par
\centerline{LAPP, IN2P3-CNRS, B. P. 110, F-74941 Annecy-le-Vieux Cedex,
France}\bigskip
 \centerline{\bf M. Fontannaz}\smallskip
\centerline{Laboratoire de Physique ${\rm Th \acute e orique}$ et Hautes
Energies\footnote{**}{Laboratoire ${\rm associ \acute e}$ au CNRS (URA
63)}}\par
\centerline{${\rm Universit \acute e}$ de Paris XI, ${\rm b \hat atiment}$ 211,
91405 Orsay Cedex, France}\par \vskip 1 truecm
\centerline{(revised, May 1994)}
\vskip 2 truecm \baselineskip=20pt
\noindent \underbar{\bf Abstract} \par
We discuss in detail the photon structure function beyond the leading logarithm
approximation. Of special concern is the factorization scheme and the hadronic
input ; we show how to naturally absorb large terms due to the
$\overline{\rm MS}$ factorization scheme in a modified hadronic component. The
effect of the charm quark mass threshold is also discussed in relation to the
phenomenology. A comparison with data shows that the modified hadronic
component
can be reasonably described by a VDM-type input. \par \vskip 2 cm
{\parskip=0cm
\noindent ENSLAPP-A-435-93 \par
\noindent LPTHE Orsay 93-37} \par
\noindent December 1993

\vfill \supereject
\noindent {\bf 1. \underbar{Introduction}} \vskip 4mm
For the past fifteen years, the photon structure function ${\rm
F}_2^{\gamma}({\rm x}, {\rm Q}^2)$ has generated considerable theoretical and
experimental work [1], and, more recently, the beginning of HERA has
reactivated the interest in the quark and gluon distributions in real photons.
\par
It is indeed expected that photoproduction experiments [2, 3] will allow to
measure these distributions with a good precision. Another advantage of
photoproduction reactions consists in the coupling of the gluon (from the
photon) to the hard subprocesses which offers a direct determination of its
distribution [4] ; in $\gamma \gamma^{\ast}$ DIS experiments it is only through
the evolution equations for ${\rm F}_2^{\gamma}$ that the gluon distribution
may
be determined. \par Jet production in $\gamma \gamma$ collisions also offers
the
possibility to measure the quark and gluon distributions in real photons, and
data from TRISTAN [5] are encouraging. The x-region probed at HERA (x $\sim$
.05) and TRISTAN (x $\sim$ .4) are complementary whereas the ${\rm
Q}^2$-regions
are very similar (${\rm Q}^2 \simeq {\rm p}_{\bot}^2 = 25 {\rm GeV}^2$).
Therefore the joint study of HERA and TRISTAN [6] results should constraint
rather tightly these parton distributions. \par
In this paper we study the problems of the factorization scheme and of the non
perturbative input in detail. In particular we show how the non perturbative
component must be modified in order to take into account the specificity
of the $\overline{\rm MS}$ factorization scheme. A new set\footnote{*}{These
distributions are obtainable from Fontanna @ qcd.th.u-psud.fr} of parton
distributions in the real photon is proposed, which incorporates this new
input. The perturbative component and the non perturbative component are
treated separately, which allows to easily modify the normalization of the non
perturbative contribution described by the Vector Meson Dominance Model (VDM).
These distributions are obtained by solving the inhomogeneous Altarelli-Parisi
(AP)
equations which incorporate beyond leading logarithm (BLL) corrections.  \par
Sections 2 and 3 discuss the factorization scheme and the non-perturbative
inputs.
The case of the massive quarks is treated in section 4 and in section 5
presents
some phenomenological applications, and a discussion of the sensitivity to the
factorization scale. Appendix A studies the new ''non perturbative`` input, and
appendix B discusses the scale and factorization problem.  \vskip 5mm
\noindent
{\bf 2. \underbar{Evolution equations and factorization scheme}} \vskip 4mm
\vskip
5mm The evolution of the gluon distribution ${\rm G}^{\gamma}({\rm x}, {\rm
Q}^2)$,
of the singlet distribution $\sum^{\gamma}({\rm x}, {\rm Q}^2) =
\displaystyle{\sum^{\rm N_f}_{\rm f=1}} ({\rm q}_{\rm f}^{\gamma}({\rm x}, {\rm
Q}^2) + \bar{\rm q}_{\rm f}^{\gamma}({\rm x}, {\rm Q}^2)) \equiv
\displaystyle{\sum_{\rm f}} {\rm q_f^{(+)}(x, Q^2)}$ and of the non-singlet
distributions ${\rm q_f^{NS}(x, Q^2)} = {\rm q_f^{(+)}} - \sum^{\gamma}/{\rm
N_f}$ (${\rm N_f}$ is the number of flavors) are governed by the inhomogeneous
Altarelli-Parisi equations [1] \vskip 4 mm $${\partial \Sigma^{\gamma} \over
\partial \log Q^2} = k_q + P_{qq} \otimes \Sigma^{\gamma} + P_{qg} \otimes
G^{\gamma} \eqno(2.1a)$$  \vskip 3 mm $${\partial G^{\gamma} \over \partial
\log
Q^2} = k_g + P_{gq} \otimes \Sigma^{\gamma} + P_{gg} \otimes G^{\gamma}
\eqno(2.1b)$$ \vskip 4 mm $${\partial q_{\rm f}^{\rm NS} \over \partial \log
Q^2} = \sigma_{f}^{NS} k_q + P_{NS} \otimes q_f^{NS} \eqno(2.2)$$   \vskip 4 mm
\noindent where $\sigma_{\rm f}^{\rm NS} = ({\rm e}_{\rm f}^2/<{\rm e}^2> -
1)/{\rm N_f}$ with $<{\rm e}^{\rm m}> = \displaystyle{\sum_{\rm f}} {\rm
e}_{\rm
f}^{\rm m}/{\rm N_f}$. The convolution $\otimes$ is defined by
\vskip 4 mm
$$P \otimes q = \int_x^1 {dz \over z} P \left ( {x \over z} \right ) q(z) \ \ \
{}.
\eqno(2.3)$$
\vskip 4 mm
The inhomogeneous $({\rm k_i})$ and homogeneous $({\rm P}_{\rm ij})$ splitting
functions have an expansion in $\alpha_{\rm s}({\rm Q}^2)$
\vskip 4 mm
$$k_q = {\alpha \over 2 \pi} k_q^{(0)} + {\alpha \over 2 \pi} {\alpha_s(Q^2)
\over 2 \pi} k_q^{(1)} \eqno(2.4)$$
\vskip 4 mm
$$k_g = {\alpha \over 2 \pi} {\alpha_s(Q^2) \over 2 \pi} k_g^{(1)} \eqno(2.5)$$
\vskip 4 mm
$$P_{ij} = {\alpha_s \over 2 \pi} P_{ij}^{(0)} + \left ( {\alpha_s \over 2 \pi}
\right )^2 P_{ij}^{(1)} \ \ \ . \eqno(2.6)$$
\vskip 4 mm
\noindent The functions ${\rm P}_{\rm ij}^{\rm (n)}$ are available in the
literature [7]. The inhomogeneous splitting functions may be derived from
the ${\rm P}_{\rm ij}$ and are given in refs. [8, 9]. \par
In terms of the parton distributions, the photon structure function is
written
$${\cal F}_2^{\gamma}(x, Q^2) \equiv F_2^{\gamma}(x, Q^2)/x = \sum_f e_f^2
\ q_f^{(+)} \otimes C_q$$
$$\hskip 3.5 truecm + \ G^{\gamma} \otimes C_g$$
$$\hskip 3 truecm + \ C_{\gamma} \ \ \ . \eqno(2.7)$$
\vskip 4 mm
\noindent The Wilson coefficients ${\rm C_q}$ and ${\rm C_g}$ may be found
in ref. [10]. The direct term ${\rm C}_{\gamma}$ is given in [11, 12]
\vskip 4 mm
$$C_{\gamma} = {\alpha \over 2 \pi} 3 \sum_f e_f^4 \ 2 \left [ \left ( x^2 + (1
-
x)^2 \right ) \ell n {1 - x \over x} + 8 x(1 - x) - 1 \right ] \ \ \ .
\eqno(2.8)$$
\vskip 4 mm
The physical quantity ${\cal F}_2^{\gamma}$ is factorization scheme
independent. This means that it does not depend on the procedure (the
factorization scheme) used to define the BLL splitting function ${\rm P}_{\rm
ij}^{\rm (n)}$ (${\rm n} \geq 1$) and ${\rm k}_{\rm i}^{\rm (n)}$, and the
function ${\rm C_q}$, ${\rm C_g}$ and ${\rm C}_{\gamma}$. This is however true
only if these functions were calculated to all orders in $\alpha_{\rm s}$. If
the truncated series (2.4) to (2.6) are used the photon structure functions is
still scheme independent, but only at order $\alpha_{\rm s}^{-1}$ and
$\alpha_{\rm s}^0$. \par In order to study this dependence in detail, let us
write explicitly the solution of eq. (2.2). Working with moments $({\rm q}_{\rm
f}^{\rm NS}({\rm n}, {\rm Q}^2) = \int_0^1 {\rm dx} \ {\rm x}^{\rm n-1} {\rm
q}_{\rm f}^{\rm NS} ({\rm x}, {\rm Q}^2))$, we get [11, 1]
\vskip 4 mm
$$q_f^{NS}(n, Q^2) = q_f^{AN}(n, Q^2) + q_f^{NP}(n, Q^2) \eqno(2.9)$$
\vskip 4 mm
\noindent with (dropping the variable n and keeping the terms of order ${\rm
O}(1/\alpha_{\rm s})$ and ${\rm O}(1)$)
$$q_f^{AN}(Q^2) = {4 \pi \over \alpha_s(Q^2)} \left [ 1 - \left (
{\alpha_s(Q^2) \over \alpha_s(Q_0^2)} \right )^{1-d} \right] a \left \{ 1 +
{\alpha_s(Q^2) \over 2 \pi} {2 \beta_1 \over \beta_0} \left ( {d \over 4} -
{P_{NS}^{(1)} \over \beta_1} \right ) \right \}$$  $$+ \left [ 1 - \left (
{\alpha_s(Q^2) \over \alpha_s(Q_0^2)} \right )^{-d} \right ] {2 \beta_1 \over
\beta_0 d} \left \{ {\alpha \over 2 \pi} \left ({k_q^{(0)} \over 2 \beta_0} -
{k_q^{(1)} \over \beta_1} \right ) \sigma_f^{NS}  + 2 \left ({d \over 4} -
{P_{NS}^{(1)} \over \beta_1} \right ) a (1 - d) \right \} \eqno(2.10)$$ \vskip
5 mm
\noindent and \vskip 4 mm
$$q_f^{NP}(Q^2) = \left ( {\alpha_s(Q^2) \over \alpha_s(Q_0^2)} \right )^{-d}
\left ( 1 + \left ( \alpha_s(Q^2) - \alpha_s(Q_0^2) \right ) R \right )
q_f^{NP}(Q_0^2) \ \ \ , \eqno (2.11)$$
\vskip 4 mm
\noindent where ${\rm a} = {\alpha \over 2 \pi \beta_0} {1 \over 1 - {\rm d}}
{\rm k}_{\rm q}^{(0)} \sigma_{\rm f}^{\rm NS}$, ${\rm R} = {\beta_1 \over \pi
\beta_0} \left ( {{\rm d} \over 4} - {{\rm P}_{\rm NS}^{(1)} \over \beta_1}
\right )$ and ${\rm d} = 2 {\rm P}_{\rm NS}^{(0)}/ \beta_0$. (The notation AN
is
for anomalous [13] and NP for nonperturbative). The running coupling constant
$\alpha_{\rm s}({\rm Q}^2)$ evolves according to
\vskip 4 mm
$${\partial \alpha_s(Q^2) \over \partial \log Q^2} = - \alpha_s \left (
{\alpha_s \over 4 \pi} \beta_0 + \left ( {\alpha_s \over 4 \pi} \right )^2
\beta_1 \right ) = \beta \left ( \alpha_s \right ) \ \ \ . \eqno(2.12)$$
\vskip 4 mm
This way of writing the solution of (2.2) emphasizes [12] the role of the
boundary condition at ${\rm Q}^2 = {\rm Q}_0^2$ where ${\rm q}_{\rm f}^{\rm
AN} ({\rm Q}^2)$ vanishes. This anomalous part of the solution (2.9) is fully
calculable within a perturbative approach [13]. The "non perturbative" part
${\rm q}_{\rm f}^{\rm NP} ({\rm Q}^2)$ is a general solution of the
homogeneous evolution equation (eq. (2.2) without the inhomogeneous term ${\rm
k}_{\rm q}$) ; it is an arbitrary function of n at ${\rm Q}^2 = {\rm Q}_0^2$,
only its evolution with ${\rm Q}^2$ is predicted. \par
We now introduce the following physical assumption. For too small values of
${\rm Q}^2$, a perturbative approach has no meaning ; the coupling constant
$\alpha_{\rm s}({\rm Q}^2)$ being too large, ${\rm q}_{\rm f}^{\rm NS}({\rm
Q}^2)$ has only a non perturbative component ${\rm q}_{\rm f}^{\rm NP}({\rm
Q}^2)$. When ${\rm Q}^2$ increases above ${\rm Q}_0^2$ a perturbative
component appears. This is our definition of ${\rm Q}_0^2$ which is no more an
arbitrary parameter, but has a well defined meaning ; it is the value of ${\rm
Q}^2$ at which the perturbative component ${\rm q}_{\rm f}^{\rm AN}({\rm
Q}^2)$ vanishes. \par
The value of ${\rm Q}_0^2$ and the n dependence of ${\rm q}_{\rm
f}^{\rm NP}({\rm Q}_0^2, {\rm n})$ is not calculable from the QCD Lagrangian
and
we must content ourselves with models. We shall examine below whether ${\rm
q}_{\rm f}^{\rm NP}$ may be described by VDM [1] when ${\rm Q}_0$ is
close the $\rho$-meson mass, ${\rm Q}_0^2 \simeq {\rm m}_{\rho}^2$. \par
Let us now come back to the problem of the factorization scheme. As explained
above, the functions ${\rm P}_{\rm ij}^{(1)}$, ${\rm k}_{\rm i}^{(1)}$ and
${\rm C}_{\rm i}$ are scheme-dependent, but not the physical quantity
${\cal F}_2^{\gamma}$. This is a well-known result for the proton structure
function [14], and here we only focus on the features of a photon target,
namely
the function ${\rm k}_{\rm q}^{(1)}$ and ${\rm C}_{\gamma}$ (for simplicity, we
only consider the part of (2.7) proportional to ${\rm e}_{\rm f}^4)$. \par
{}From (2.10) we see that ${\rm q}_{\rm f}$ changes when we modify the
factorization scheme. But ${\rm C}_{\gamma}$ is also modified so that the sum
\vskip 4 mm
$$- {\alpha \over 2 \pi} {k_q^{(1)} \over P_{qq}^{(0)}} {<e^4> \over <e^2>} +
C_{\gamma} \eqno(2.13)$$
\vskip 4 mm
\noindent which appears in (2.7) is factorization scheme independent [15]. \par
Let us now assume that we work in a given factorization scheme called "tilde"
to which correspond the functions $\tilde{k}^{(1)}_q$ and $\tilde{C}_{\gamma}$.
We get for ${\cal F}_2^{\gamma}$ in this new scheme, keeping only the terms
which interest us
\vskip 4 mm
$${\cal F}_2^{\gamma} = \cdots - {\alpha \over 2 \pi} \left [ 1 - \left (
{\alpha_s(Q^2) \over \alpha_s(Q_0^2)} \right )^{-d} \right ] {<e^4> \over
<e^2>} {\tilde{k}_q^{(1)} \over P_{qq}^{(0)}} + \cdots + \tilde{C}_{\gamma}$$
$$\hskip -3 truecm + \left ( {\alpha_s(Q^2) \over \alpha_s(Q_0^2)}
\right )^{-d} \sum_{f=1}^{N_f} e_f^2 \ \tilde{q}_f^{NP}(Q_0^2) \ \ \
,\eqno(2.14)$$ \vskip 4 mm
\noindent expression that we compare with that obtained in the
$\overline{\rm MS}$ factorization scheme [16], in which we have the functions
$k_q^{(1)}$ and $C_{\gamma}$
\vskip 4 mm
$${\cal F}_2^{\gamma} = \cdots - {\alpha \over 2 \pi} \left [ 1 - \left (
{\alpha_s(Q^2) \over \alpha_s(Q_0^2)} \right )^{-d} \right ] {<e^4> \over
<e^2>} \ {k_q^{(1)} \over P_{qq}^{(0)}} + \cdots + C_{\gamma}$$
$$\hskip -3 truecm + \left ( {\alpha_s(Q^2) \over \alpha_s(Q_0^2)}
\right)^{-d} \sum_f e_f^2 \ q_f^{NP}(Q_0^2) \ \ \ .\eqno(2.15)$$
\vskip 4 mm
Assuming that we know the relation between the two schemes
\vskip 4 mm
$$\tilde{k}_q^{(1)} = k_q^{(1)} + \delta k_q^{(1)} \ \ \ , \eqno(2.16)$$
\vskip 4 mm
\noindent we must have for consistency
\vskip 4 mm
$$C_{\gamma} = \tilde{C}_{\gamma} - {\alpha \over 2 \pi} {<e^4> \over <e^2>}
{\delta k^{(1)} \over P_{qq}^{(0)}} \eqno(2.17)$$
\vskip 4 mm
and
\vskip 4 mm
$$q_f^{NP} = \tilde{q}_f^{NP} + {\alpha \over 2 \pi} \ {e_f^2 \over
N_f<e^2>} \ {\delta k_q^{(1)} \over P_{qq}^{(0)}} \ \ \ . \eqno(2.18)$$
\vskip 4 mm
{}From expression (2.18) we explicitly verify that the ``non perturbative''
input
is scheme dependent, and contains a perturbative part ! Thus the assumption
that
the ``non perturbative'' input could be described by a VDM-type input is
clearly
too naive [17]. It may however be true in a specific factorization scheme and
we
explore this possibility in section 3.
 \par \vskip 5 mm

\noindent {\bf 3.
\underbar{The non perturbative input and the Vector Dominance Model}}   \vskip
4mm
In order to better understand the result of the preceding section, let us
consider
the lowest order contribution to ${\cal F}_2^{\gamma}$ coming from the
imaginary
part of the Box diagram, fig. 1, which shows how the virtual photon q probes
the
quark content of the real photon p. The lower blob ${\rm G}({\rm k}^2)/{\rm
k}^2$
(it includes the quark propagators) represents the pointlike coupling of the
real
photon to a ${\rm q}\bar{\rm q}$ pair and all the non perturbative effects as
sketched in fig. 2. \par Actually our only assumption is that ${\rm G}({\rm
k}^2)/{\rm k}^2$ tends to the pointlike term for large $|{\rm k}^2|$.

$$\lim_{|{\rm k}^2| >> \Lambda^2} G(k^2) = 1 \ \ \ . \eqno(3.1)$$
\vskip 4 mm
\noindent When ${\rm k}^2$ goes to zero, ${\rm G}({\rm k}^2)/{\rm k}^2$ must
be integrable, because ${\cal F}_2^{\gamma}$ is a finite physical
quantity. This means that we must have $\displaystyle{\lim_{{\rm k}^2 \to 0}}
{\rm G}({\rm k}^2) \sim (- {\rm k}^2)^{\alpha}$ with $\alpha > 0$. \par
We make the pointlike content of ${\rm G}({\rm k}^2)$ explicit by writing
\vskip 4 mm
$${G(k^2) \over k^2} = {1 \over k^2} + {G(k^2) - \theta(|k^2| - Q_0^2) \over
k^2} - {\theta (Q_0^2 - |k^2|) \over k^2} \ \ \ . \eqno(3.2)$$ \vskip 4 mm
\noindent The first term on the RHS of (3.2), without cut on ${\rm k}^2$,
corresponds to the perturbative expression of the Box diagram. \par
Its contribution, in the collinear approximation, is easily calculated [4].
Introducing the Sudakov decomposition
\vskip 4 mm
$$k = zp + {k^2 + \vec k_{\bot}^2 \over z 2n.p} n + k_{\bot} \eqno (3.3)$$
\vskip 4 mm
\noindent with ${\rm k}_{\bot}.{\rm p} = {\rm k}_{\bot}.{\rm n} = 0$,
$\vec{\rm k}_{\bot}^2 = - {\rm k}_{\bot}^2$, n.p large, and working in $4 -
2\varepsilon$ dimensions, we have (dropping all inessential factors)
\vskip 4 mm
$${\cal F}_2^{\gamma} \sim \int {dz \over z} \int_0^{q^2/x} dk^2 \int d
\vec{k}_{\bot}^2 \left ( \vec{k}_{\bot}^2 \right )^{- \varepsilon} \delta
\left ( \vec{k}_{\bot}^2 + k^2(1 - z) \right ) \delta \left ( {z \over x} - 1
+ (1 - x) {k^2 \over Q^2} \right ) |M|^2  \eqno(3.4)$$
\vskip 4 mm
\noindent where ${\rm Q}^2 = - {\rm q}^2$ and ${\rm x} = {\rm Q}^2/2 {\rm
p.q}$.
The matrix element squared is (in the collinear approximation
and in a physical gauge) given by (for one quark species)
\vskip 4 mm
$$|M|^2 \sim e_f^4 \ {\alpha \over 2 \pi} \ {(\mu^2)^{\varepsilon} \over k^2} \
{3 \over 1 - \varepsilon} \left ( z^2 + (1 - z)^2 - \varepsilon \right ) \ \ \
{}.
\eqno(3.5)$$  \vskip 4 mm  \noindent Keeping only the terms coming from the
collinear $1/{\varepsilon}$ pole, we get (adding quark and antiquark
contributions, and putting ${1 \over \bar{\varepsilon}} = {1 \over \varepsilon}
- \gamma_E + \ell n \ 4 \pi$) \vskip 4 mm  $${\cal F}_2^{\gamma}(x) \sim 3
e_f^4 \
{\alpha \over \pi} \left [ (x^2 + (1 - x)^2) \left ( - \ {1 \over
\bar{\varepsilon}} \left ( {Q^2 \over \mu^2} \right )^{- \varepsilon} \right )
+
\left ( x^2 + (1 - x)^2 \right ) \ell n {1 - x \over x} + 2x(1 - x) \right ]$$
$$\equiv 3 e_f^4 \ {\alpha \over \pi} \left ( - {1 \over \bar{\varepsilon}} +
\ell
n {Q^2 \over \mu^2} \right ) \left ( x^2 + (1 - x)^2 \right ) + C_{\gamma ,
c}^f \
\ \ . \eqno(3.6)$$  \vskip 4 mm  This expression for the box diagram has been
obtained with the dimensional regularization which is the one used to define
the
$\overline{\rm MS}$ factorization scheme. Here it consists in subtracting [16]
the
term proportional to $({\rm Q}^2 / \mu^2)^{- \varepsilon}/\bar{\varepsilon}$.
This
procedure defines the (scheme dependent) direct term ${\rm C}_{\gamma , {\rm
c}}^{\rm f}$ (or ${\rm C}_{\gamma}$ given in (2.8) when the full calculation of
the Box diagram is performed). \par
${\cal F}_2^{\gamma}({\rm x})$ being a physical quantity, it cannot
contain the $1/\varepsilon$ pole, and it is here that the third term of the RHS
of (3.2) plays its role. We obtain from this last term
\vskip 4 mm
$${\cal F}_2^{\gamma, 3} \sim - 3 e_f^4 \ {\alpha \over \pi} \left \{
\left ( - {1 \over \bar{\varepsilon}} + \ell n {Q_0^2 \over \mu^2} \right )
\left
(x^2 + (1 - x)^2 \right ) + \left ( x^2 + (1 - x)^2 \right ) \ell n (1 - x) +
2x(1 - x) \right \}$$
$$\equiv - 3 e_f^4 \ {\alpha \over \pi} \left ( - {1 \over \bar{\varepsilon}} +
\ell n {Q_0^2 \over \mu^2} \right ) \left ( x^2 + (1 - x)^2 \right ) -
C_{\gamma , 3}^f \ \ \ .   \eqno(3.7)$$
\vskip 4 mm
This term has no anomalous $\ell n {\rm Q}^2$ behavior. Actually it is
independent of ${\rm Q}^2$, when QCD is not switched on. (We postpone to
appendix A a discussion of the effects of perturbative QCD corrections). When
the
part of ${\cal F}_2^{\gamma_{,3}}$ proportional to $- {1 \over
\bar{\varepsilon}} +
\ell n {{\rm Q}_0^2 \over \mu^2}$ is added to (3.6), the $1/\varepsilon$ poles
cancel each other and $\ell n {\rm Q}^2/\mu^2$ is changed into  \vskip 4 mm
$$\ell n {Q^2 \over Q_0^2} = \ell n {Q^2 \over \Lambda^2} - \ell n {Q_0^2 \over
\Lambda^2} = {4 \pi \over \beta_0 \alpha_s(Q^2)} \left ( 1 - \left (
{\alpha_s(Q^2) \over \alpha_s(Q_0^2)} \right ) \right ) \ \ \ . \eqno(3.8)$$
\vskip 4 mm \noindent This ${\rm Q}^2$-dependence corresponds to the LL part of
(2.10) with d = 0. \par Let us now consider the second term of (3.2). The
$\theta$-function cuts the $1/{\rm k}^2$ perturbative behavior of this
contribution. The integration over ${\rm k}^2$ is therefore controlled by the
non perturbative behavior of ${\rm G}({\rm k}^2)/{\rm k}^2$ and we obtain a
result which does not depend on ${\rm Q}^2$. Figure 3 shows the ${\rm k}^2$
behavior of this term. The value of ${\rm Q}_0^2$ must of course be chosen such
that ${\rm G}^{\rm H}({\rm k}^2, {\rm Q}_0^2) \equiv {\rm G}({\rm k}^2) -
\theta
\left ( |{\rm k}^2| - {\rm Q}_{0}^2 \right )$ represents the non perturbative
input. For instance a too large value of
${\rm Q}_0^2$ would include in ${\rm G}^{\rm H}({\rm k}^2)$ a perturbative
tail.
It is reasonable to assume that ${\rm G}^{\rm H}({\rm k}^2, {\rm Q}_0^2)$ can
be
approximately described by VDM with ${\rm Q}_0^2 \simeq {\rm m}_{\rho}^2$ and
we
write (for ${\rm Q}^2 >> \Lambda^2$) \vskip 4 mm $$\int_0^{Q^2} {dk^2 \over
k^2}
G^H(k^2, m_{\rho}^2) \simeq \int_0^{\infty} {dk^2 \over k^2} G^H(k^2,
m_{\rho}^2)
\equiv q^{VDM}(Q^2 = m_{\rho}^2) \ \ \ .\eqno(3.9)$$
\vskip 4 mm
\noindent With (3.9) we have defined a non perturbative input (if $Q_0^2$ is
correctly chosen) which is {\bf invariant with respect to the photonic
factorization scheme}. (This point is fully developed in Appendix B). When the
QCD
evolution is switched on, both ${\rm q}^{\rm VDM}$ and ${\rm C}_{\gamma , 3}$
acquire a hadronic ${\rm Q}^2$-dependence and we get (Appendix A)
\vskip 4 mm $$q_f^{NP}(Q^2) = \left ( {\alpha_s(Q^2) \over
\alpha_s(Q_0^2)} \right )^{-d} \left ( q_f^{VDM}(Q_0^2) - {C_{\gamma , 3}^f
\over
e_f^2} \right ) \ \ \ . \eqno(3.10)$$ \vskip 4 mm
\noindent Therefore, in the $\overline{\rm MS}$ factorization scheme, the "non
perturbative" input, that we call $\overline{\rm VDM}$, is given by expression
(3.10), and, at ${\rm Q}^2 = {\rm Q}_0^2$, we have
\vskip 4 mm
$${\cal F}_2^{\gamma}(x, Q_0^2) = C_{\gamma}(x) + \sum_{f=1}^{N_f}
e_f^2 \ q_f^{VDM}(Q_0^2) - C_{\gamma , 3} \ \ \ . \eqno(3.11)$$
\vskip 4 mm
In the above expression, we have only studied the part of ${\cal
F}_2^{\gamma}$ associated to the quark contributions. Similar
considerations show that the gluonic contribution is given by VDM only. These
results are very similar to those obtained by the authors of ref. [9] who
worked
in a different factorization scheme called ${\rm DIS}_{\gamma}$ in which
\vskip
4 mm $${\alpha \over 2 \pi} \ {<e^4> \over <e^2>} k_q^{(1)}(DIS_{\gamma}) =
{\alpha \over 2 \pi} \ {<e^4> \over <e^2>} \ k_q^{(1)}(\overline{MS}) -
C_{\gamma} P_{qq}^{(0)} \ \ \ , \eqno(3.12)$$ \vskip 4 mm \noindent so that
$C_{\gamma}(DIS_{\gamma}) = 0$. \par  They then \underbar{assume} [18] that, in
this
scheme, the non perturbative input is given by VDM \vskip 4 mm
$$q_f^{NP}{(Q^2)_{DIS}}_{\gamma} = q_f^{VDM}(Q^2) \ \ \ , \eqno(3.13)$$
\vskip 4 mm
\noindent and get
\vskip 4 mm
$${\cal F}_2^{\gamma}(x, Q_0^2) = \sum_f e_f^2 q_f^{VDM}(Q_0^2) \ \ \ .
\eqno(3.14)$$ \vskip 4 mm
\noindent Expressed in our language (and in the $\overline{\rm MS}$ scheme),
the assumption of the authors of ref. [9] means that ${\rm q}_{\rm f}^{\rm NP}$
is given by (3.10), but with ${\cal C}_{\gamma , 3}$ replaced by ${\rm
C}_{\gamma}$. However the expression ${\rm C}_{\gamma}$ is process dependent
(here the $\gamma \gamma^{\ast}{\rm DIS}$) ; it includes for instance $\ell n \
{\rm x}$ which comes from the upper limit of the ${\rm k}^2$-integration in
(3.4). On the contrary our derivation of (3.10) with $C_{\gamma_{,3}}^f$, is
process independent (all process dependent terms are eliminated by the pole
approximation). Our definition of the quark distributions in the photon is
therefore universal, and the evolutions are performed in the $\overline{MS}$
scheme. \par Before giving some numerical results showing the difference
between
the naive input (${\rm q}^{\rm NP} = {\rm q}^{\rm VDM}$) and the $\overline{\rm
VDM}$ input (3.10), let us make some more precise statements on VDM. \par We
consider that the photon is a coherent superposition of vector mesons  \vskip 4
mm $$\gamma = {g \over \sqrt{2}} \left ( \rho + {\omega \over 3} - {\sqrt{2}
\over 3} \phi \right ) = g \left ( {2 \over 3} u \bar u - {1 \over 3} d \bar d
-
{1 \over 3} s \bar s \right ) \eqno(3.15)$$ \vskip 4 mm \noindent with a
coupling constant g determined from  \vskip 4 mm $$\sigma_{tot}(\gamma p) =
{g^2
\over 2} \left ( \sigma_{tot}( \rho p) + {1 \over 9} \sigma_{tot}(\omega p) +
{2
\over 9} \ \sigma_{tot} (\phi p) \right )$$  $$\simeq {g^2 \over 2} \ {12 \over
9} \ \sigma_{tot}(\rho p) \simeq g^2 \ {2 \over 3} \  \sigma_{tot}(\pi p) \ \ \
. \eqno (3.16)$$ \vskip 4 mm \noindent A comparison between the data on the
total cross-sections of photon-proton and proton-proton reactions leads to
\vskip 4 mm $$g^2 \simeq \alpha \ \ \ . \eqno(3.17)$$ \vskip 4 mm
\noindent Assuming that the parton distributions in the ${\rm q} \bar{\rm q}$
''bound states'' of (2.27) are similar to those of the pion, observed in
Drell-Yan and direct photon reactions [19], we can write
\vskip 4 mm
$$u_{valence}^{\gamma}(x, Q^2) = g^2 \ {4 \over 9} \ u_{valence}^{\pi}
\eqno(3.18a)$$
\vskip 4 mm
$$u_{sea}^{\gamma}(X, Q^2) = g^2 \left ( {4 \over 9} + {1 \over
9} + {1 \over 9} \right ) u_{sea}^{\pi} = g^2 \ {2 \over 3} \
u_{sea}^{\pi} \eqno(3.18b)$$
\vskip 4 mm
$$G^{\gamma}(x, Q^2) = g^2 {2 \over 3} G^{\pi}(x, Q^2) \eqno(3.18c)$$
$$\hskip -3 truecm \vdots$$
\vskip 4 mm
\noindent and so on for the parton distributions of the VDM component of the
real photon. \par
This model is rough and it must be considered as a ''zero order'' approximation
of the non perturbative part of the photon structure function, which should be
determined from data [2, 3, 5]. In the parametrization of the distribution
functions presented in this paper, we leave an arbitrary factor in front of the
VDM contribution so that the user can modify the normalization of this input.
\par
We are now ready to compare the naive input (expression (3.10) with
$C_{\gamma_{,3}}^f = 0$) with the $\overline{\rm VDM}$ input. The theoretical
calculations are compared at ${\rm Q}^2 = 73 \ {\rm GeV}^2$ with data from AMY
[20] in the Fig. 4. For these predictions, we chose ${\rm Q}_0^2 = .25 \ {\rm
GeV}^2$ and took into account the mass of the charm quark following the
approach
described in the next section. We see that the difference between the two
inputs
is small, the $\overline{\rm VDM}$ input improving slightly the agreement with
data. Notice that at very large ${\rm Q}^2$, the difference between the two
inputs must disappear, because both have a hadronic ${\rm Q}^2$-behaviour which
vanishes for asymptotic values of ${\rm Q}^2$. We also show in this figure the
result obtained (with the $\overline{\rm VDM}$ input) when we choose ${\rm
Q}_0^2 = 1 \ {\rm GeV}^2$. The shorter ${\rm Q}^2$-evolution of the anomalous
component explains the difference between the full and dotted curves.
 \vskip 5mm
\noindent {\bf 4. \underbar{Treatment of the massive quarks}}    \vskip 4mm
Threshold effects due to the charm quark may be important in ${\cal
 F}_2^{\gamma}(x, Q^2)$ at large x. To study this problem, let us again
consider
the Box diagram and the massive quark contribution to ${\cal F}_2^{\gamma}$.
Dropping all inessential factors, one gets [13, 21] \vskip 4 mm $${\cal
F}_{2,c}^{\gamma}(x, Q^2) \sim \theta(1 - \beta) \left \{ \left ( x^2 + (1 -
x)^2
+ {4 m_c^2 \over Q^2} x(1 - 3x) - 8 {m_c^4 \over Q^4} x^2 \right ) \ell n
{\left
( 1 + \sqrt{1 - \beta} \right )^2 \over \beta} \right .$$  $$\hskip 4.5 truecm
\left . + \left ( 8 x(1 - x) - 1 - {4 m_c^2 \over Q^2} x(1 - x) \right )
\sqrt{1
- \beta} \right \} \eqno(4.1)$$  \vskip 4 mm \noindent with $\beta = 4 {\rm
m}_{\rm c}^2/{\rm s}$ and ${\rm s} = ({\rm p} + {\rm q})^2$. By neglecting the
higher twist terms proportional to ${\rm m}_c^2 /{\rm Q}^2$, we can rewrite
(4.1) $${\cal F}_{2,c}^{\gamma} \sim \theta (1 - \beta ) \left \{ \left ( x^2 +
(1 - x)^2 \right ) \ell n {Q^2 \over m_c^2} + \left ( x^2 + (1 - x)^2 \right )
\ell n \left ( \left ( {1 + \sqrt{1 - \beta} \over 2} \right )^2 {s \over Q^2}
\right ) \right .$$   $$\hskip 5 truecm \left . + \left ( 8 x (1 - x) - 1
\right
) \sqrt{1 - \beta} \right \} \ \ \ . \eqno(4.2)$$
\vskip 4 mm
\noindent Subtracting the term proportional to $\log {{\rm Q}^2 \over {\rm
m}_{\rm c}^2}$, we define a massive direct term
$$C_{\gamma}(x, m_c) = e_c^4 {\alpha \over \pi} 3 \theta(1 - \beta) \left [
\left
( x^2 + (1 - x)^2 \right ) \ell n \left ( \left ( {1 - x \over x} \right )
\left ( {1 + \sqrt{1 - \beta} \over 2} \right )^2 \right ) \right .$$ $$\left .
\hskip 3 truecm + \left ( 8 x (1 - x) - 1 \right ) \sqrt{1 - \beta} \right ]
\eqno(4.3)$$
\noindent which has the massless limit (2.8) when $\beta = 0$. \par
The term proportional to $\ell n {{\rm Q}^2 \over {\rm m}_{\rm c}^2}$ is
generated by the massless evolution equations (2.1) and (2.2) with the
boundary condition ${\rm q}_{\rm c}({\rm x}, {\rm Q}^2 = {\rm m}_{\rm c}^2) =
0$. This evolution exactly reproduces, at the lowest order in $\alpha_{\rm s}$,
the $\ell n {{\rm Q}^2 \over {\rm m}_{\rm c}^2}$-term of (4.2). In the full
solution (2.7, 2.10) however, our procedure does not lead to a complete
description of the threshold behavior, because logarithmic terms similar to
that of (4.3) should also be included in ${\rm k}_{\rm q}^{(1)}$, ${\rm
C}_{\rm q}$, and a two loop calculation of ${\rm C}_{\gamma}(x, {\rm m})$
should be performed [22]. \par
The effect of the massive direct term (4.3) is important when x goes to
${\rm x}_{\rm th} = 1 /(1 + 4 {\rm m}_{\rm c}^2 / {\rm Q}^2)$. One then get
$\ell n {(1 + \sqrt{1 - \beta} )^2 \over \beta} \sim 2 \sqrt{1 - \beta}$ in
(4.1) and ${\cal F}_{2,c}^{\gamma}$ given by (4.2) goes to zero, whereas
the use of the massless limit (2.8) (without cut on x) leads to a negative
contribution when x goes to 1. The difference between the massless and the
massive approach is shown in fig. 5. It is non negligible for ${\rm x} \geq
{\rm x}_{\rm th}$. For larger values of ${\rm Q}^2$, ${\rm x}_{\rm th}$ is
closer
to 1.0 and beyond the accessible experimental $x$-range. \par One must however
keep in mind that in most applications [4, 6], we are far from the threshold
and
the massless evolution of the charm distribution is a good approximation which
allow to take into account the effects of the QCD evolution, not present in
(4.2). The distributions presented in this paper are obtained by solving eq.
(2.1) and (2.2) with ${\rm N_f} = 3$ for ${\rm Q}_0^2 \leq {\rm Q}^2 \leq {\rm
m}_{\rm c}^2$ and ${\rm N_f} = 4$ for ${\rm Q}^2 > {\rm m}_{\rm c}^2$. (${\rm
m}_{\rm c}^2 = 2 {\rm GeV}^2$). Therefore both the anomalous and the
$\overline{\rm VDM}$ components have no charm content for ${\rm Q}^2 \leq {\rm
m}_{\rm c}^2$. \vskip 5mm
\noindent {\bf 5. \underbar{Phenomenological analysis}}
\vskip 4mm
In this section we give some numerical examples of the distributions
described in this paper and we compare our predictions for ${\cal
F}_2^{\gamma}$
with data. On this occasion, we also discuss the respective sizes of the
anomalous and VDM components, the constraints coming from experimental
results, and the sensitivity to the factorization scale. \par Let us start with
our standard set of distributions defined as follows. Each distribution
consists
of an anomalous part, which vanishes at ${\rm Q}^2 = {\rm Q}_0^2$, and a "non
perturbative" part given by $\overline{\rm VDM}$ (3.10). The VDM contribution
of
(3.10) is explicitly described in (3.18). A free normalization factor K is put
in front of this contribution. In the standard set, K = 1 and ${\rm Q}_0^2 = .5
\ {\rm GeV}^2$. The distributions are solutions of the BLL (in the
$\overline{\rm MS}$ scheme) Altarelli-Parisi equations (2.1) and (2.2) with
$\Lambda_{\overline{\rm MS}} = .200 {\rm GeV}$ (the sensitivity to the value of
$\Lambda_{\overline{\rm MS}}$ is very small). The charm quark distributions
vanishes at ${\rm Q}_0^2 = {\rm m}_{\rm c}^2 = 2 \ {\rm GeV}^2$. We do not
compare the LL distributions with the BLL ones, this point being already
discussed in ref. [23]. \par
The singlet $\sum^{\gamma}({\rm x}, {\rm Q}^2)$ and gluon ${\rm
G}^{\gamma}({\rm x}, {\rm Q}^2)$ distributions are shown in figs. 6 and 7 for
${\rm Q}^2 = 45 \ {\rm GeV}^2$, a typical value observed in the experiments.
The
crosses are the full distributions and the dots correspond to a
VDM input set equal to zero (K = 0). We see that the VDM contribution is
important at small x and negligible at large x (the behavior of
$\sum^{\gamma}({\rm x}, {\rm Q}^2)$ at large x is due to the $\ell {\rm n}^2(1
-
{\rm x})$ term of the BLL splitting function ${\rm k}_{\rm q}^{(1)}$).
Therefore, depending on the x-domain, it is possible to explore either the
perturbative or the non perturbative component of the distributions. For
instance the photoproduction of large-${\rm p}_{\bot}$ jets or particles
at HERA [4, 6, 23] essentially explores the small x domain of the
distributions, sensitive to the VDM input (more precisely, to the gluon
content of the VDM input). \par
In DIS $\gamma \gamma^{\ast}$ experiments, a wider x-domain is accessible,
as shown in figs. 4, 8, 9, 10 and 11 where predictions are compared with data.
We also show in figs. 8, 9, 10 and 11 the results obtained with K = 0. The
agreement with data is good for $Q^2 = 4.2 \ {\rm GeV^2}$ and satisfactory for
larger values of $Q^2$. The accuracy of PLUTO data at $Q^2 = 4.2 \ {\rm GeV^2}$
put
stringent constraints on the VDM component which appears to have the right
normalization. A more important VDM component is suggested by the data at
larger
$Q^2$~; they are however not very accurate and we cannot draw any precise
conclusion from them. We could also vary the value of $Q_0^2$ in order to fit
the
experimental results. We see from figures 8 and 9, corresponding to large
values
of $Q^2$ and $x$ where the VDM component is small, that a small value of
$Q_0^2$ is favored. (The variations of $F_2^{\gamma}$ with $Q_0^2$ have been
discusssed in section 3 and shown in fig. 4).  \par

Until now the factorization scale ${\rm M}^2$ has been taken equal to ${\rm
Q}^2$.
The more general case is treated in appendix B, and we give here some numerical
results. We found that ${\cal F}_2^{\gamma}$ is very stable with respect to
variation of the factorization scale ${\rm M}^2$. For instance at ${\rm Q}^2 =
73 \
{\rm GeV}^2$, for $.25 \leq x \leq .8$, ${\cal F}_2^{\gamma}$ decreases by less
than $6 \ \%$ when ${\rm M}^2$ changes from ${\rm Q}^2/4$ to $4 {\rm Q}^2$. At
the
same time, the singlet distribution increases by some $50 \  \%$. Therefore we
cannot blame the scale dependence for a possible disagreement between theory
and
data. \par

Two other BLL parametrizations of the quark and gluon distributions in the
photon have already been published. In the one by Gordon and Storrow [27], the
input distribution at $Q_0^2 = 5.3 \ {\rm GeV^2}$ contains, besides the VDM
contribution, a box-diagram component involving the quark masses. The free
parameters are then fixed by a fit to PLUTO data [25] at $Q_0^2$. Data at
higher
values of $Q^2$ are well described by this model. The assumptions involved in
the Gl\"uck-Reya-Vogt parametrization [18] are very close to those done in this
paper ; the main difference, besides a different value of $Q_0^2$ and of the
normalization of the VDM component, is the one indicated by eqs. 3.11 and 3.14,
and already discussed in section 3. Readers interested by numerical comparisons
will find in ref. [28] a critical discussion of the three parametrizations.
\par

Let us end this section by a remark on the jet production in $\gamma
\gamma$ collision at TRISTAN [5] where the two almost real photons are defined
by
an anti-tag condition. The measured cross-section ${\rm d} \sigma ({\rm e}^+
{\rm
e}^- \to {\rm jet + X})/{\rm dp}_{\bot}$ corresponds to the convolution of the
$\gamma \gamma$ cross-section with the ${\rm Weizs \ddot a cker}$-Williams
distributions of photons in the electron. Because of this convolution, the
${\rm
e}^+{\rm e}^-$ cross-section is not sensitive to the shape of the VDM quark and
gluon distributions ; the detailed x-behavior of these distributions is washed
out
by the convolution [6]. As a result the cross-section ${\rm d} \sigma({\rm
e}^+{\rm
e}^-)/{\rm d}{\rm p}_{\bot}$ of these ''anti-tagged'' experiments are only
sensitive to the overall normalization of the VDM component. \vskip 5 mm

\vfill \supereject
\centerline{\bf \underbar{Conclusion}} \vskip 4
mm In this paper we have proposed a set of parton distributions in the real
photon
which incorporates a modified VDM input, called $\overline{VDM}$, and in this
way
takes into account the specificity of the $\overline{MS}$-factorization scheme.
\par The agreement between theory and data is satisfactory in what regards the
photon structure function. The large error bars do not allow to draw any
definite
conclusion and other comparisons with experiment are necessary. \par At the
prospect of the study of other reactions involving the parton distribution in
the
photon, we leave free the normalization of the VDM input so that it can be
adjusted
to get a better agreement with future data. We would like however to dwell upon
the
fact that VDM is only a model and that there exist other predictions concerning
the
non perturbative input [34]. In fact this latter must be determined from data
and
this should be one of the main goals of photoproduction experiments.

\vfill \supereject
\centerline{\bf \underbar{Appendix A}} \par  \bigskip We
elaborate on the ''cut`` Box contribution $\theta(Q_0^2 - |k^2|)/k^2$ discussed
in
section 3 at the lowest order in the strong coupling constant. At all orders in
$\alpha_s$, we get a generalized ladder [29,4] with an integration on the
virtuality ${\rm k}^2$ of the quark emitted by the real photon bounded by $-
{\rm
Q}_0^2$. This all-order expression may be split in various contributions (we
consider the part of $q_f^{NS}$ proportional to $e_f^2$ and drop all
unnecessary
indices) :\par \hskip 1 cm 1) a contribution \vskip 4 mm $$q_f(M^2, Q_0^2) =
{e^2
\over N_f<e^2>} \int_0^{-Q_0^2} {dk^2 \over k^2} k_q(k^2, \varepsilon) \
e^{\int_{k^2}^{-M^2} {dk'^2 \over k'^2} P(k'^2, \varepsilon)} \eqno(A.1)$$
\vskip
4 mm \noindent which regularizes the singular quark distribution $q_f(M^2,
M^2)$
such that \vskip 4 mm
$$q_f(M^2) = q_f(M^2, M^2) - q_f(M^2, Q_0^2) = {e^2 \over N_f<e^2>}
\int_{-Q_0^2}^{-M^2} {dk^2 \over k^2} k_q \ e^{\int_{k^2}^{-M^2} {dk'^2 \over
k'^2} P} \eqno(A.2)$$
\vskip 4 mm
\noindent has no pole in $1/\varepsilon$ and vanishes at $M^2 = Q_0^2$, \par
\hskip 1 cm 2) a ''cut`` direct contribution $C_{\gamma , 3}^f(\alpha_s)$,
already calculated at order $\alpha_s^0$ in section 3, \par
\hskip 1 cm 3) a dressed ''cut`` Box contribution
\vskip 4 mm
$$\delta q_f(M^2) = \int_{-Q_0^2}^{-M^2} {dk^2 \over k^2} \delta k_q \
e^{\int_{k^2}^{-M^2} {dk'^2 \over k'^2} P} \eqno(A.3)$$ \vskip 4 mm
\noindent with, at order $\alpha_s$,
\vskip 4 mm
$$\delta k_q = {\alpha_s \over 2 \pi} P^{(0)} C_{\gamma , 3}^f/e_f^2 = {\alpha
\over 2 \pi} \ {\alpha_s \over 2 \pi} \delta k_q^{(1)} \ \ \ . \eqno(A.4)$$
\vskip 4 mm
\noindent Therefore ${\cal F}_2^{\gamma}$ can be written (for one flavour and
forgetting $q_f^{VDM}$) $${\cal F}_2^{\gamma} = e_q^2 \ C_q \ q_f(Q^2) +
C_{\gamma}^f - \left ( e_q^2 \ C_q \ \delta q_f(Q^2) + C_{\gamma , 3}^f \right
) \
\ \ . \eqno(A.5)$$ \vskip 4 mm
\noindent In this expression, which is very similar to the one advocated by
the authors of ref. [9], the ''cut`` contributions must not change the
asymptotic $Q^2$-behavior. To see how the cancellations occur, let us write
\vskip 4 mm
$$\delta q_f = \delta q_f^U + \delta q_f^D \eqno(A.6)$$
\noindent with
$$\delta q_f^U(M^2) = \int_0^{\alpha_s(M^2)} {d \alpha '_s \delta k_q(\alpha
'_s)
\over \beta(\alpha '_s)} e^{\int_{\alpha '_s}^{\alpha_s(M^2)} {d \alpha ''_s
\over
\beta(\alpha ''_s)} P(\alpha ''_s)} \eqno(A.7)$$ \vskip 4 mm \noindent
\noindent and
$$\eqalignno{
\delta q_f^D(M^2) &= - e^{\int_{\alpha_s(Q_0^2)}^{\alpha_s(M^2)} {d \alpha ''
\over \beta} P} \int_0^{\alpha_s(Q_0^2)} {d \alpha ' \delta k_q \over
\beta} e^{\int_{\alpha '_s}^{\alpha_s(Q_0^2)} {d \alpha ''
\over \beta} P} \cr
&= - e^{\int_{\alpha_s(Q_0^2)}^{\alpha_s(M^2)}} \delta q^U(Q_0^2) \ \ \ .
&(A.8)}$$ \noindent
The expression $e_q^2 \ C_q \ \delta q_f^U(Q^2)$, which can be expanded in
power of
$\alpha_s(Q^2)$, cancels the direct contribution $C_{\gamma , 3}^f
(\alpha_s(Q^2))$ (this is trivially verified at order $\alpha_s^0$), so
that the ''cut`` contribution to ${\cal F}_2^{\gamma}$ (besides A.1) is
just $- e_q^2 \ C_q \ \delta q_f^D(Q^2)$ which has an hadronic behaviour in
$\left
( \alpha_s(Q^2)/\alpha_s(Q_0^2) \right )^{-2 P^{(0)}/\beta_0}$. This last
contribution, with $e_q^2 \ \delta q_f^D(Q_0^2) = C_{\gamma , 3}^f$ (expression
3.7), is the one we have introduced in our definition of $\overline{VDM}$
in section 3. \par
Because the same factorization procedure is at work in the calculation of
$q_f$ and $\delta q_f$, the difference $q_f-\delta q_f$ is factorization
scheme invariant, as far as the  splitting function $k_q$ is
concerned. The same remark is true for $C_{\gamma}^f - C_{\gamma ,
3}^f$.

\vfill \supereject
\centerline{\bf \underbar{Appendix B}} \par
\bigskip
In order to study the scale and factorization scheme independence of ${\cal
F}_2^{\gamma}$, let us start from the all order expression (dropping
inessential charge factors) :\par  \vskip 4 mm
$${\cal F}_2^{\gamma} = C_q(M) \left \{
\int_{\alpha_s(Q_0)}^{\alpha_s(M)} {d \alpha '_s \ k_q(\alpha '_s)
\over \beta(\alpha '_s)} e^{\int_{\alpha '_s}^{\alpha_s(M)} {d \alpha
''_s \ P(\alpha ''_s) \over \beta(\alpha ''_s)}} + q^{NP}(M) \right \}
+ C_{\gamma}(M) \eqno(B.1)$$ \vskip 4 mm
\noindent where $k_q$, $P$, $\beta$ are series in $\alpha_s$, and $C_q(M)$ and
$C_{\gamma}(M)$ are series in $\alpha_s(\mu)$ ($\mu$ is the renormalization
scale) ; $q^{NP}(M)$ is a "non perturbative" input. By introducing a variation
$\delta k_q$ of $k_q$ (of order $0(\alpha_s^n)$, $n \geq 1$), we get~:

\vskip 4 mm
$${\cal F}_2^{\gamma} = C_q(M) \left \{
\int_{\alpha_s(Q_0)}^{\alpha_s(M)} d \alpha '_s \ {k_q
+ \delta k_q \over \beta} e^{\int_{\alpha
'_s}^{\alpha_s(M)} d \alpha ''_s {P\over \beta}} + q^{NP}(M) +
\int_0^{\alpha_s(Q_0^2)} d \alpha '_s {\delta k_q \over \beta} \right .$$
$$\left . e^{\int_{\alpha '_s}^{\alpha_s(M)} d \alpha ''_s {P \over \beta}}
\right
\} + C_{\gamma}(M) - C_q(M) \int_0^{\alpha_s(M)} d \alpha '_s {\delta k_q \over
\beta} e^{\int_{\alpha '_s}^{\alpha_s(M)} d \alpha ''_s {P \over \beta} } \ \ \
.\eqno(B.2)$$ \vskip 4 mm We see that both the "non perturbative" input and the
direct term are modified by the variation of the splitting function $k_q$. \par
A variation $\delta P$ of $P$ leads to
\vskip 4 mm
$${\cal F}_{\gamma}^2 = \left ( C_q(M) \ e^{- \int_0^{\alpha_s(M)} {d \alpha
'_s
\ \delta P \over \beta}} \right ) \left \{ \int_{\alpha_s(Q_0)}^{\alpha_s(M)}
{d
\alpha '_s \over \beta} \left (  k_q \ e^{\int_0^{\alpha '_s}
d \alpha ''_s {\delta P \over \beta}} \right )   \right .$$ $$\left .
e^{\int_{\alpha '_s}^{\alpha_s(M)} d \alpha ''_s {P + \delta P \over \beta}} +
e^{\int_{\alpha_s(Q_0)}^{\alpha_s(M)} d \alpha '_s {P + \delta P \over \beta}}
\left ( e^{\int_0^{\alpha_s(Q_0^2)} d \alpha '_s {\delta P \over \beta}}
q^{NP}(Q_0) \right ) \right \}$$  $$\hskip - 5 truecm + C_{\gamma}(M) \ \ \ ,
\eqno(B.3)$$ \vskip 4 mm  \noindent which shows that $\delta P$ modifies the
inhomogeneous splitting function $k_q$, the ''non perturbative`` input at $M =
Q_0$, and the Wilson coefficient $C_q(M)$. \par The invariance of ${\cal
F}_2^{\gamma}$ with respect to variations of $\mu$ and $M$ can be studied in a
similar way. The invariance of ${\cal F}_2^{\gamma}$ is not true anymore when
the calculation is performed with truncated series. In this case we can use
Stevenson's approach [30, 31, 32] to find an optimum of ${\cal F}_2^{\gamma}$
with
respect to $\mu$, $M$, $P$ and $k$. We must however notice that there is a
mixing
between the ''non perturbative`` contribution and the anomalous contribution
(for
the part proportional to $(\alpha_s(M)/\alpha_s(Q_0))^{-2P^{(0)}/\beta_0})$, as
shown in eq. (B.2), so that the non perturbative input cannot be factored out
the
optimization equation, as in the hadronic case [31]. \par To better delimit the
scheme dependence of the non perturbative input and to make the connection with
section 3, we expand the part of ${\cal F}_2^{\gamma}$ proportional to
$(\alpha_s(M)/\alpha_s(Q_0)^{-2P^{(0)}/\beta_0}$ in power of $\alpha_s$
\vskip 4 mm
$$C_q(M) \left ( q^D(M) + q^{NP}(M) \right ) = C_q(M) \left (
\alpha_s(M)/\alpha_s(Q_0) \right )^{-2P^{(0)}/\beta_0} {\alpha \over 2 \pi}
\left (
1 - {\alpha_s(M) \over 2 \pi} {2P^{(1)} \over \beta_0} \right ) .$$
$$\left ( {2 \pi \over \alpha_s(Q_0)} \ {2 k_q^{(0)} \over 2 P^{(0)} - \beta_0}
+
{k_q^{(1)} \over P^{(0)}} + {k_q^{(0)} \over P^{(0)}} \ {2 P^{(1)} \over
\beta_0}
+ {2 \pi \over \alpha} q^{NP} (Q_0) \right ) \ \ \ . \eqno(B.4)$$
\vskip 4 mm
\noindent We have neglected terms of order $0(\alpha_s(Q_0))$ in the last
bracket. Expression (B.4) shows that
\vskip 4 mm
$${k_q^{(1)} \over P^{(0)}}&Q + {k_q^{(0)} \over P^{(0)}} \ {2P^{(1)} \over
\beta_0} + {2 \pi \over \alpha} \ q^{NP}(Q_0) = \kappa^{NP} \eqno(B.5)$$ \vskip
4
mm \noindent must be an invariant of the factorization schemes. From (B.3) we
see
that a variation of $P$ causes a variation of $k_q$ so that ${k_q^{(1)} \over
P^{(0)}} + {k_q^{(0)} \over P^{(0)}} \ {2P^{(1)} \over \beta_0}$ is an
invariant ; therefore $q^{NP}(Q_0)$ only depends on the factorization scheme
defining $k_q^{(1)}$. In the approach discussed in section 3, we put
\vskip 2mm
$$q^{NP}(Q_0) = q^{VDM}(Q_0) \ - {\alpha \over 2 \pi} \ {\delta k_q^{(1)} \over
P^{(0)}} \eqno(B.6)$$ \vskip 4 mm
\noindent or, through (B.5)
\vskip 3 mm
$${k_q^{(1)} - \delta k_q^{(1)} \over P^{(0)}} + {k_q^{(0)} \over P^{(0)}} \
{2P^{(1)} \over \beta_0} + {2 \pi \over \alpha} q^{VDM}(Q_0) = \kappa^{NP}
\eqno(B.7)$$ \vskip 4 mm \noindent which makes evident the invariance of
$q^{VDM}(Q_0)$ with respect to the factorization scheme defining $k_q$.
(Another
way to introduce a ''non perturbative`` component and to regularize the
solutions of (2.1) and (2.2) is given in [33]).

\par In this paper, we are interested only in the variation of ${\cal
F}_2^{\gamma}$ with $M$, the factorization scale (setting $\mu = M$). By
imposing
the consistency condition that the variation of ${\cal F}_2^{\gamma}$ with $M$
vanishes at the lowest order in $\alpha_s$, we get the following relations for
$C_q(M) = 1 + {\alpha_s(M) \over 2 \pi} C_q^{(1)}(M)$, $C_g(M) = {\alpha_s(M)
\over 2 \pi} C_g^{(1)}(M)$ and $C_{\gamma}(M)$~:

\vskip 2mm
$$\eqalignno{
& {\partial C_q^{(1)} \over \partial \log M^2} = - P_{qq}^{(0)} = - C_F \left (
{1 + x^2 \over 1 - x} \right )_+ &(B.8) \cr
&{\partial C_g^{(1)} \over \partial \log M^2} = - <e^2> P_{qg}^{(0)} = - 2 N_f
<e^2> {1 \over 2} \left [ x^2 + (1 - x)^2 \right ] &(B.9) \cr
&{\partial C_{\gamma} \over \partial \log M^2} = - 2 N_f <e^4> {\alpha \over 2
\pi} k_q^{(0)} = - {\alpha \over 2 \pi} 2N_f <e^4> 3 \left [ x^2 + (1 - x)^2
\right ] \ \ \ . &(B.10) \cr
}$$
\vskip 3 mm

Therefore, when the factorization scale $M$ is different from $\sqrt{Q^2}$, we
have

\vskip 3mm
$$C_{\gamma}(M) = {\alpha \over 2 \pi} 2 N_f<e^4> 3 \left [ x^2 + (1 - x)^2
\right ] \log {Q^2 \over M^2} + C_{\gamma} \ \ \ ,  \eqno(B.11)$$
\vskip 3 mm
\noindent and similar expressions for $C_q^{(1)}(M)$ and $C_g^{(1)}(M)$.

\vfill \supereject
\centerline{\bf \underbar{References}} \par
\vskip 5mm
\item{[1]} Reviews on the photon structure function may be found
in C. Berger and W. Wagner, Phys. Rep. $\underline{146}$ (1987) 1 ; H.
Kolanoski and P. Zerwas, in High Energy ${\rm e}^+$-${\rm e}^-$ physics, World
Scientific, Singapore, 1988, Eds. A. Ali and P. ${\rm S\ddot o ding}$. \par
\item{} J. H. Da Luz Vieira and J. K. Storrow, Z. Phys. $\underline{C51}$
(1991) 241. \par
\item{[2]} H1 Collaboration, T. Ahmed et al., Phys. Lett. $\underline{B297}$
(1992) 205.\par
\item{[3]} Zeus Collaboration, M. Derrick et al., Phys. Lett.
$\underline{B297}$ (1992) 404. \par
\item{[4]} For a review on the photoproduction at HERA, see M. Fontannaz,
"The Photon Structure Function at HERA", Orsay preprint LPTHE 93-22, talk
given at the XXI International Meeting on Fundamental Physics, Miraflores
de la Sierra, Spain (May 1993).\par
\item{[5]} TOPAZ Collaboration, H. Hayashii et al., Phys. Lett.
$\underline{B314}$
(1993) 149. \par \item{} AMY Collaboration, R. Tanaka et al., Phys. Lett.
$\underline{B277}$ (1992) 215. \par   \item{[6]} P. Aurenche, M. Fontannaz, J.
Ph.
Guillet, Y. Shimizu, J. Fujimoto and K. Kato, KEK preprint 93-180 (Dec. 1993).
\par   \item{[7]} W. Furmanski and R. Petronzio, Phys. Lett. $\underline{B97}$
(1980) 435. \par    \item{[8]} M. Fontannaz and E. Pilon, Phys. Rev.
$\underline{D45}$ (1992) 382. \par    \item{[9]} M. ${\rm Gl \ddot u ck}$, E.
Reya
and A. Vogt, Phys. Rev. $\underline{D45}$ (1992) 3986. \par \item{[10]} W. A.
Bardeen, A. J. Buras, D. W. Duke and T. Muta, Phys. Rev. $\underline{D18}$
(1978)
3998. \par \item{} G. Altarelli, R. K. Ellis and G. Martinelli, Nucl. Phys.
$\underline{B157}$ (1979) 461. \par   \item{[11]} W. A. Bardeen and A. J.
Buras,
Phys. Rev. $\underline{D20}$ (1979) 166 ; $\underline{21}$ (1980) 2041 (E).
\par
\item{[12]} M. ${\rm Gl \ddot u ck}$ and E. Reya, Phys. Rev. $\underline{D28}$
(1983) 2749. \par
\item{[13]} E. Witten, Nucl. Phys. $\underline{120}$ (1977) 189. \par
\item{[14]} A. J. Buras, Rev. Mod. Phys. $\underline{52}$ (1980) 199. \par
\item{[15]} T. Uematsu and T. F. Walsh, Nucl. Phys. $\underline{B199}$ (1982)
93. \par
\item{[16]} W. Furmanski and R. Petronzio, Z. Phys. $\underline{C11}$ (1982)
293.  \par
\item{[17]} A critical discussion of the VDM assumption has been given by J. A.
Da Luz Vieira and J. K. Storrow (ref. [1]) in a different context. \par
\item{[18]} M. Gl\"uck, E. Reya and A. Vogt, Phys. Rev. $\underline{D46}$
(1992)
1973. \par
\item{[19]} P. Aurenche, R. Baier, M. Fontannaz, M. H.
Kienzle-Focacci and M. Werlen, Phys. Lett. $\underline{B233}$ (1989) 517. \par
\item{[20]} T. Sasaki et al., Phys. Lett. $\underline{B252}$ (1990) 491. \par
\item{[21]} C. T. Hill and G. C. Ross, Nucl. Phys. $\underline{148}$ (1979)
373.
\par \item{} M. ${\rm Gl \ddot u ck}$ and E. Reya, Phys. Lett. \underbar{83B}
(1979) 98. \par  \item{[22]} E. Laenen, S. Riemersma, J. Smith and W.L. van
Neerven, Fermilab-Pub-93/240-T. \par
\item{[23]} P. Aurenche, P. Chiappetta, M. Fontannaz, J. Ph. Guillet et E.
Pilon,
Z. Phys. $\underline{C56}$ (1992) 589. \par     \item{[24]} JADE
Collaboration, W. Bartel et al., Z. Phys. $\underline{C24}$ (1984) 231. \par
\item{[25]} PLUTO Collaboration, Ch. Berger
et al., Nucl. Phys. $\underline{B281}$ (1987) 365 ; Phys. Lett.
$\underline{142B}$
(1984) 11. \par
\item{[26]} TASSO Collaboration, M. Althoff et al.,  Z. Phys.
$\underline{C31}$ (1986) 527.
\item{[27]} L. E. Gordon and J. K. Storrow, Z. Phys. $\underline{C56}$ (1992)
307.
\par
\item{[28]} A. Vogt, Proceedings of the Workshop ``Two-photon physics at LEP
and
HERA'', Lund, May 1994. \par
 \item{[29]} G. Curci, W. Furmanski and R. Petronzio,
Nucl. Phys. $\underline{B175}$ (1980) 27. \item{[30]} P. M. Stevenson, Phys.
Rev.
$\underline{D23}$ (1981) 2916. \item{[31]} P. M. Stevenson and H. D. Politzer,
Nucl. Phys. $\underline{B277}$ (1986) 758.
\item{[32]} H. Nakkagawa, A. Niegawa and H. Yokota, Phys. Rev.
$\underline{D33}$
(1986) 46.
\item{[33]} I. Antoniadis et G. Grunberg, Nucl. Phys. $\underline{B213}$ (1983)
455.
\item{[34]} A. S. Gorski, B. L. Ioffe, A. Yu. Khodjaminian, A. Oganesian, Z.
Phys.
$\underline{C44}$ (1989) 523.
  \vfill \supereject
\centerline{\bf \underbar{Figure Captions}} \par
\vskip 5mm
{\parindent=2cm
\item{\bf Fig. 1  :} The Box diagram.
\item{\bf Fig. 2  :} The perturbative and non perturbative content
of the lower blob of fig. 1.
\item{\bf Fig. 3  :} The behavior with ${\rm k}^2$ of ${\rm G}^{\rm
H}({\rm k}^2, {\rm Q}_0^2)$.
\item{\bf Fig. 4  :} Predictions for ${\rm F}_2^{\gamma}$ compared
with AMY data [20]. Naive VDM input and ${\rm Q}_0^2 = .25 \ {\rm GeV}^2$
(dashed
curve) ; $\overline{\rm VDM}$ input and ${\rm Q}_0^2 = .25 \ {\rm GeV}$ (full
curve)~; $\overline{\rm VDM}$ input and ${\rm Q}_0^2 = 1. \ {\rm GeV}^2$
(dotted
curve).
\item{\bf Fig. 5  :} The massive (dots) and massless (crosses) photon structure
function at ${\rm Q}^2 = 9.2 \ {\rm GeV}^2$.  \item{\bf Fig. 6  :} The singlet
distribution at ${\rm Q}^2 = 45 \ {\rm GeV}^2$ (crosses). With K = 0 (dots).
\item{\bf Fig. 7  :} The gluon distribution ; same conventions as in fig. 6.
\item {\bf Fig. 8  :} JADE data [24]. The standard prediction
(see text) (full curve). With K = 0 (dashed curve).
\item{\bf Fig. 9  :} PLUTO data [25]. Same as fig. 8.
\item{\bf Fig. 10 :} TASSO data [26].
Same as fig. 9.  \par
\item{\bf Fig. 11 :} PLUTO data [25]. Same as fig. 8. \par
}
 \bye